# BACOM 2.0 facilitates absolute normalization and quantification of somatic copy number alterations in heterogeneous tumor

Yi Fu[1], Guoqiang Yu[1], Douglas A. Levine[2], Niya Wang[1], Ie-Ming Shih[3], Zhen Zhang[3], Robert Clarke[4] and Yue Wang[1]

[1]Department of Electrical and Computer Engineering, Virginia Polytechnic Institute and State University, Arlington, VA 22203, USA; [2]Department of Surgery, Memorial Sloan-Kettering Cancer Center, New York, NY 10021, USA; [3]Departments of Pathology and Oncology, Johns Hopkins University, Baltimore, MD 21231, USA; [4]Lombardi Comprehensive Cancer Center, Georgetown University, Washington, DC 20057, USA

## 1 INTRODUCTION

In the November 2011 issue, Yu *et al.* proposed a Bayesian analysis of copy number mixtures (BACOM) method to detect genomic deletion types and to correct normal cell contamination in copy number data (Yu, et al., 2011). They tested BACOM method on two simulated and two prostate cancer datasets, and obtained very promising results supported by the ground truth and biological plausibility (Zhang, et al., 2014). In a subsequent analysis of TCGA ovarian cancer dataset, the average normal cell fraction estimated by BACOM was found higher than expected. In this letter, we first discuss the advantages of BACOM method in relation to alternative approaches (Carter, et al., 2012; Rasmussen, et al., 2011; Su, et al., 2012; Yuan, et al., 2012). Then, we show that this elevated estimate of normal cell fraction is the combined result of incorrect signal normalization and parameter estimation. Lastly, we describe an allele-specific absolute normalization and quantification scheme that can enhance BACOM applications in many biological contexts (Kuhn, et al., 2012; Liu, et al., 2009). An open-source MATLAB software is developed to implement BACOM 2.0 and is publically available.

## 2 EXTENDED METHOD

### 2.1 BACOM overview

BACOM is a statistically principled and unsupervised method that detects copy number deletion types (homozygous versus heterozygous), estimates normal cell fraction, and recovers cancer-specific copy number profiles, using allele-specific copy number signals. In a heterogeneous tumor sample, the measured copy number intensity is a mixture of the signals from both normal and cancer cells,

$$X_i = \alpha \times X_{\text{normal},i} + (1-\alpha) \times X_{\text{cancer},i}, \quad (1)$$

where $X_i$ is the observed copy number signal at locus $i$, $\alpha$ is the unknown fraction of normal cells, $X_{\text{normal},i}$ and $X_{\text{cancer},i}$ are the latent copy number signals in normal and cancer cells at locus $i$. Let $X_{A,i}$ and $X_{B,i}$ be the allele-specific copy number signals, $X_i = X_{A,i} + X_{B,i}$ are assumed to be independently and identically distributed random variables following a normal distribution $\mathcal{N}(\mu_{A+B}, \sigma^2_{A+B})$ whose mean $\mu_{A+B}$ and variance $\sigma^2_{A+B}$ can be readily estimated by the sample averages. Allele-specific analyses are focused on the deletion regions with distinct genotypes.

Types of deletions are detected by a model-based Bayesian hypothesis testing. Specifically, BACOM uses a novel summary statistic,

$$Y = \sigma^{-2}_{A-B} \sum_{i=1}^{L} (X_{A,i} - X_{B,i})^2, \quad (2)$$

where $\sigma^2_{A-B}$ is the variance of $X_{A,i} - X_{B,i}$ in a length-$L$ deletion region. It has been shown that under homo-deletion, $Y$ follows an $L$ degrees of freedom standard $\chi^2$ distribution, given by

$$\chi^2(y;L) = \begin{cases} \dfrac{1}{2^{L/2}\Gamma(L/2)} y^{(L/2)-1} e^{-y/2} & \text{for } y > 0, \\ 0 & \text{for } y \le 0, \end{cases} \quad (3)$$

and under hemi-deletion, $Y$ follows an $L$ degrees of freedom noncentral $\chi^2$ distribution, given by

$$\chi^2(y;L,\lambda) = \begin{cases} \dfrac{e^{-(y+\lambda)/2}}{2^{L/2}} \displaystyle\sum_{k=0}^{\infty} \dfrac{y^{L/2+k-1}\lambda^k}{\Gamma(k+L/2)2^{2k}k!} & \text{for } y > 0, \\ 0 & \text{for } y \le 0, \end{cases} \quad (4)$$

where $\lambda = L(2 - \mu_{A+B})^2 \sigma^{-2}_{A+B}(1+\rho)/(1-\rho)$, $\rho$ is the genuine correlation coefficient between $X_{A,i}$ and $X_{B,i}$, and $\Gamma$ denotes the Gamma function. Since for a deletion region, we have

$$\begin{cases} E[X_i] = \alpha \times 2 + (1-\alpha) \times 0 = 2\alpha, & \text{if homo-deletion,} \\ E[X_i] = \alpha \times 2 + (1-\alpha) \times 1 = 1+\alpha, & \text{if hemi-deletion,} \end{cases} \quad (5)$$

then, the average normal cell fraction $\overline{\alpha}$ across the whole genome can be estimated, as well as cancer-specific copy number profiles, given by

$$\hat{X}_{\text{cancer},i} = \frac{X_i - 2\overline{\alpha}}{1 - \overline{\alpha}}, \text{ with } \alpha_{\text{homo}} = \frac{E[X_i]}{2}, \alpha_{\text{hemi}} = E[X_i] - 1, \quad (6)$$

### 2.2 Problem diagnosis

In our independent analyses of TCGA samples with BACOM, we confirmed unexpectedly higher average normal cell fractions. By a closer check on the interim results of the entire BACOM analytic pipeline, we found that many normal/amplified copy regions and hemi-deletions were misclassified as homo-deletions. This observation explains well the suspected overestimation of normal cell fraction, since $\alpha$ will be overestimated when non-deletion regions are wrongly used in (6), or $\alpha_{\text{homo}}$ is applied to hemi-deletions ($\alpha_{\text{homo}} > \alpha_{\text{hemi}}$). We thus argue that this elevated estimate is the combined result of incorrect signal modeling and normalization, particularly in the presence of copy-neutral loss of heterozygosity (LOH) and aneuploidy. For example, if a non-deletion region is firstly misclassified as deletion due to incorrect signal normalization, it can be further misclassified as homo-deletion in the cases of allelic balance. Moreover, if the value of $\rho$ is firstly underestimated due to copy-neutral LOH (allelic-imbalance) contamination in normal/allelic-balanced regions, hemi-deletion will then be misclassified as homo-deletion caused by much reduced signal-to-noise ratio.

More detailed reasoning on the root causes of the underestimated tumor purity by BACOM method are given in Supplementary Information.

## 2.3 BACOM 2.0: Allele-specific absolute normalization and quantification

Accurate signal normalization essentially rescales the relative signal intensities on the basis of normal copy regions (diploid reference loci), here termed as absolute normalization (Attiyeh, et al., 2009; Popova, et al., 2009). As the intertwined result of normal cell contamination, copy number aberrations, and tumor aneuploidy, the average ploidy of tumor cells cannot be assumed to be 2N or integer (Rasmussen, et al., 2011). Though absolute normalization is critical to inferring absolute copy numbers in a tumor sample, the classic normalization procedure based on median-centering of the total probe intensities is problematic (Carter, et al., 2012; Wang, et al., 2002; Yu, et al., 2011), since the dominant component of the intensity mixture distribution rarely coincides with the normal copy number '2' (Rasmussen, et al., 2011).

Let us consider histogram modeling of genome-wide copy number signals. Based on underlying signal characteristics, we adopt a mixture of $K$ Gaussian distributions (Attiyeh, et al., 2009), given by

$$f(x) = \sum_{k=1}^{K} \pi_k g\left(x \middle| \mu_k, \sigma_k^2\right), \tag{7}$$

where $\pi_k$ is the relative proportion of the $k$-th copy number component and $g(.|.)$ is the Gaussian kernel with $\left(\mu_k, \sigma_k^2\right)$ being the mean and variance. Such mixtures can be estimated from observed histogram using soft clustering or maximum likelihood method. However, our experimental studies on real tumor data confirmed that the component mean with the largest $\pi_k$ does not always correspond to the mean of normal copy regions, probably due to aforementioned factors, and thus cannot serve as the baseline for absolute normalization. While we have also observed that the largest component(s) often resides within the neighborhood of normal copy component.

Thus, we first develop an effective scheme to eliminate the loci belonging to the hemi-deletions (with copy number '1') and the allelic-imbalanced regions (with copy number '3', '5', etc.). Specifically, we use a sliding window centered at a locus to estimate the between-allele correlation coefficient and remove those loci whose correlation coefficients are lower than an automatically-determined threshold value, since the imbalanced allele signals associated with odd copy numbers would produce a sufficiently negative value of $\rho$, given by (in the case of copy number '3')

$$\rho_{\text{allelic-imbalanced}} \cong \frac{4(1+\rho)\sigma^2}{(1-\alpha)^2 + 4\sigma^2} - 1, \tag{8}$$

where $\sigma^2$ is the variance of noise and $\rho_0$ is the genuine between-allele correlation coefficient. It is worth noting that this procedure also eliminates copy-neutral LOH loci and thus can improve the accuracy of estimating $\rho$ by using only normal copy loci. It can be shown that, copy-neutral LOH contamination will result in an inaccurate estimate of $\rho$, given by

$$\rho_{\text{LOH-contaminated}} \cong \frac{(1+\rho)\sigma^2}{\eta(1-\alpha)^2 + \sigma^2} - 1, \tag{9}$$

where $\eta$ is the percentage of copy-neutral LOH contamination.

Subsequently, a revised Gaussian mixture model (7) is derived from solely the remaining allelic-balanced loci. Tested on many real copy number datasets, we found that the dominant component of the revised Gaussian mixture distribution now corresponds to the normal copy number regions in majority of cancer types. We thus propose to rescale the measured copy number signal intensities using the mode of the dominant component. Since such signal normalization is performed in each individual sample and based on the signals of normal copy number regions, BACOM 2.0 implements an accurate and absolute normalization (Attiyeh, et al., 2009).

Moreover, BACOM 2.0 includes an accurate estimation of allelic correlation coefficient $\rho$ (related to model parameter $\lambda$ in defining hemi-deletion summery statistic) that was often underestimated due to copy-neutral LOH contamination. Again, by excluding copy-neutral LOH loci and identifying dominant normal copy regions via aforementioned scheme, we can now obtain a more accurate estimate of allelic correlation coefficient $\rho$ and subsequently differentiate between hemi- and homo- deletions.

Also, we made an effort to calibrate allele signal crosstalk and saturation effects. Theoretically, signal crosstalk from the probes that differ only in one SNP adds positive bias to the copy number estimate that could lead to an overestimation of normal cell fraction by (6). As aforementioned, allelic crosstalk effect also bias the estimate of allele correlation coefficient. Concerning copy number signal saturation using SNP arrays, we adopted a similar linearization strategy used by ABSOLUTE (Carter, et al., 2012).

Lastly, we exploited a mathematically-justified scheme to correct the confounding impact of intratumor heterogeneity on estimating tumor purity (Oesper, et al., 2013; Rasmussen, et al., 2011). Though normal fraction $\alpha$ can hypothetically be estimated using any deletion segments, it can be experimentally and theoretically shown that the value of $\alpha$ will highly likely be overestimated when intratumor heterogeneity occurs in the deletion segment being used. Thus, in the presence of suspected intratumor heterogeneity, only the 'pure' deletion segments with homogeneous tumor genotypes should be used to estimate the normal fraction. Based on the distribution of $\alpha$ estimates across the whole genome, BACOM 2.0 calculate the final value of normal fraction using the 9-percentile of $\alpha$ estimates.

More information on BACOM 2.0 method and algorithm, summarized in Fig. 1, is included in Supplementary Information.

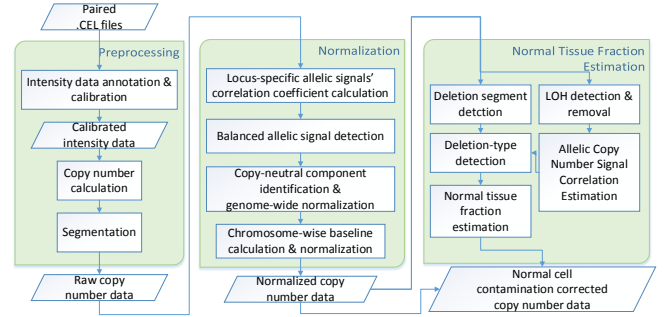

**Figure 1.** Analytic pipeline of BACOM 2.0: schematic flowchart.

In relation to previous work, the concept of using allele-specific information for analyzing copy number data is shared by others (Yau, et al., 2010), and was further developed by Rasmussen *et al.* (Rasmussen, et al., 2011) for exploratory data visualization in conjunction with visual inspection on aneuploidy and tumor heterogeneity. There is also some similarity between our objectives and others in cancer copy number restoration and tumor purity estimation. The major limitations of the approach by Yuan *et al.* (Yuan, et al., 2012) are that it requires matched genomic and histopathological image data and heavily relies on the quality of images (coarse H&E staining, artifacts, batch effects). ABSOLUTE developed by Carter et al. (Carter, et al., 2012) is supported by an elegant yet complex mathematical framework and can select the most likely combination of estimated tumor purity and ploidy by integrating copy number data and supervised learning. It is acknowledged that the cornerstone system of equations is underdetermined and various heuristics cannot guarantee a unique and correct solution (Oesper, et al., 2013). For example, in the presence of intratumor heterogeneity, the restored copy number signals are not necessarily all integer values, thus using the highest likelihood of producing all integer signals to select the most likely solution may be problematic (Oesper, et al., 2013). PurityEst proposed by Su et al. (Su, et al., 2012) estimates normal cell fraction using single-nucleotide variants but not original sequence reads. The formulation does not explicitly consider effects of copy number gains/losses thus may bias tumor purity estimation. Moreover, PurirryEst (Su, et al., 2012), THetA (Oesper, et al., 2013), and AbsCN-seq (Bao, et al., 2014) rely on next-generation sequencing data, thus may not be applicable to existing copy number data acquired using more classic methods.

# 3 EXPERIMENTAL RESULTS

## 3.1 Validation on realistic simulations

We first considered numerical mixtures of simulated normal and cancer copy number profiles across a chromosome region, a situation in which all factors are known and linear mixture model (1) is valid. We reconstituted mixed copy number signals by multiplying the simulated cancer copy number profile by the tumor purity percentage in a given heterogeneous sample. The realistic simulations were generated using a specifically selected pair of matched tumor-normal ovarian cancer samples in TCGA, where the tumor somatic copy number profile is approximately normal, i.e., allelic-balanced, summed copy number '2', and no LOH contamination. After variably dividing the whole region into eight segments, we assigned allelic-specific copy number status to each of the segments ranging from 0 to 3, as specified in Fig. 2. The raw copy number signals (the sum of the two alleles) were produced by mixing $1-\alpha$ fraction of simulated tumor copy number profile with $\alpha$ fraction of normal copy number profile, as given in (1). This simulation represents a highly challenging scenario in which the majority of probe sets were not 'normal' but amplified, yet also contained hemi-deletion segment and copy-neutral LOH segment.

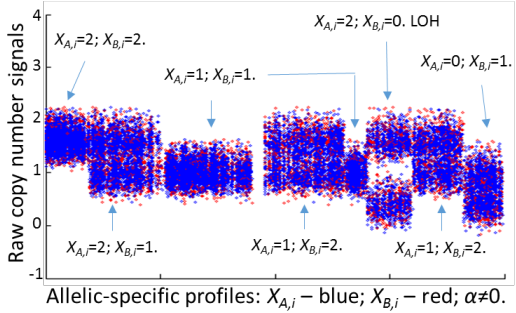

**Figure 2.** Realistic simulated allelic-specific copy number signals.

Using BACOM 2.0 analytic pipeline, we first calculated the histogram of the raw copy number signals (Fig. 3a); then we preprocessed the raw copy number signals by a moving-average low-pass filter that significantly reduced the noise effect, and re-calculated the histogram (Fig. 3b); lastly we eliminated all allelic-imbalanced loci and generated a revised histogram whose dominant peak correctly coincided with the normal copy number '2' component (Fig. 3c).

**Table 1.** Comparative parameter estimates by BACOM and BACOM 2.0

| Parameter | Ground truth | BACOM | BACOM 2.0 |
|---|---|---|---|
| $\rho$ | -0.042 | -0.714 | -0.063 |
| $\alpha$ | 40% | 79% | 39% |

With a successful absolute normalization, we first checked the estimated value of between-allele correlation coefficient $\rho$, and then recalculated the normal cell fraction $\alpha$. Based on the comparative estimates given in Table 1, the power of BACOM 2.0 is evident since the model parameter estimates were very close to the ground truth as compared to what obtained using original BACOM.

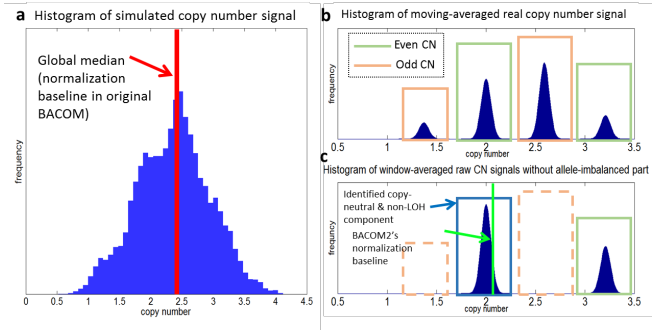

**Figure 3.** (a) Histogram of simulated copy number signals; (b) Histogram of preprocessed copy number signals after moving-average; (c) 'revised' histogram of copy numbers after eliminating allelic-imbalanced regions.

More information on validation design and experimental results is included in the supplementary information.

## 3.2 Analysis of benchmark real copy number data

We then applied BACOM 2.0 to the challenging case of TCGA ovarian cancer dataset (466 samples), where a high genomic instability has been well-documented in high-grade ovarian cancers (Kuhn, et al., 2012; Kuo, et al., 2009; Kuo, et al., 2010). We have observed that, in a large number of tumor samples, the dominant component of raw measured copy number histogram does not correspond to the normal copy number '2' but rather the allele-imbalanced loci (Fig. 4a). This observation suggests the widely existed partial aneuploidy in these samples, and the improper use of global mean/median as the normalization baseline (Attiyeh, et al., 2009).

Using BACOM 2.0 analytic pipeline, we preprocessed the raw measured copy number signals by a moving-average low-pass filter, eliminated all allelic-imbalanced loci, generated a revised histogram, and identified the component of normal copy number '2' (Fig. 4b). With a successful absolute normalization, we estimated tumor purity and tumor-specific copy number profile on each sample. From a comparison between the histogram of tumor purities likely underestimated by original BACOM (Fig. 4c) and the histogram of tumor purities newly estimated by BACOM 2.0 (Fig. 4d), we can see that BACOM 2.0 has now produced much higher tumor purity estimates (average purity of 64% versus 33%) that are theoretically expected and consistent with the protocol baseline adopted in independent studies (using 50% purity as the threshold to differentiate between high and low tumor purity in three cancer types) (Downey, et al., 2014; Huijbers, et al., 2013; Su, et al., 2012).

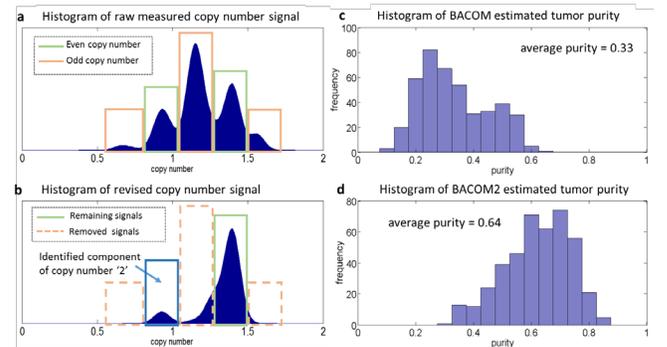

**Figure 4.** Analysis by BACOM 2.0 on the real TCGA ovarian cancer samples. (a) Histogram of copy number signals after moving-average preprocessing; (b) Histogram of 'revised' copy number signals after eliminating

allelic-imbalanced loci; (c) Histogram of tumor purity estimated by original BACOM; (d) histogram of tumor purity estimated by BACOM 2.0.

We further compared the estimates by BACOM 2.0 with the estimates by ABSOLUTE (Carter, et al., 2012) on the same datasets. As a closely relevant method, ABSOLUTE reports the estimates of tumor purity and average ploidy on two TCGA datasets, ovarian cancer (OV) and brain cancer (GBM). With a quality control selection on paired tumor and normal samples, ABSOLUTE analyzed 392 tumor samples in the OV dataset. The average tumor purity estimates by BACOM 2.0 and ABSOLUTE are 64% and 78%, respectively; and the average tumor ploidy estimates by BACOM 2.0 and ABSOLUTE are 2.33 and 2.73, respectively. The sample-wise correlation coefficients show that both tumor purity and tumor ploidy estimates by BACOM 2.0 correlate well with the estimates by ABSOLUTE (Fig. 5), achieving a high correlation coefficient of $r = 0.74$ on purity and a high correlation coefficient of $r = 0.71$ on ploidy. On the GBM dataset, the average tumor purity estimates by BACOM 2.0 and ABSOLUTE are 59% and 71%, respectively; and the average tumor ploidy estimates by BACOM 2.0 and ABSOLUTE are 2.09 and 2.17, respectively.

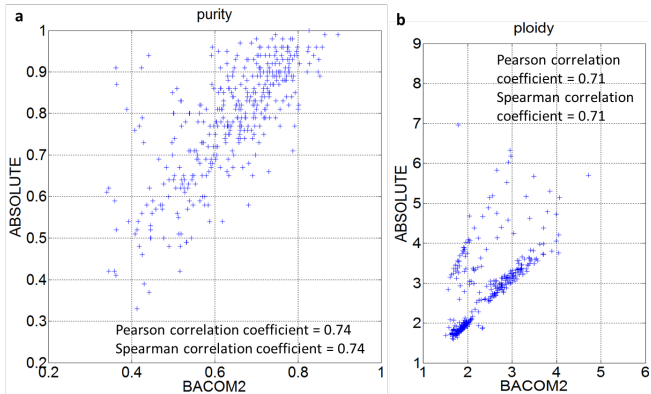

**Figure 5.** Sample-wise comparison between the estimates of tumor purity and average ploidy by BACOM 2.0 and ABSOLUTE on TCGA ovarian cancer samples. (a) Scatter plot of tumor purity estimates; (b) Scatter plot of tumor ploidy estimates.

More information on additional experimental results (tables and figures) is included in the supplementary information.

### 3.3 Cross-affirmation by expression deconvolution

In the absence of definite ground truth about the tumor purities in real samples, the validation of a new method for quantifying absolute copy numbers is always problematic. A reasonable alternative is to perform some form of 'cross' affirmation by exploiting the 'orthogonal' information structures provided by the independent sources related to a common set of nature states (Niv Ahituv and Ronen, 1988). We lastly compared the tumor purity estimates by BACOM 2.0 with the estimates by an independent method (called UNDO) that deconvoluted the mixed gene expression profiles of tumor and stroma cells acquired from the same TCGA OV samples (Wang, et al., 2013). Using the UNDO software, we analyzed the tumor samples with consistent purity estimates by both BACOM 2.0 and ABSOLUTE. The experimental result shows that the tumor purity estimates by BACOM 2.0 (based on copy number data) correlates well with the estimates by UNDO (based on gene expression data), consistently achieving a strong average 'cross' correlation coefficient of 0.5~0.6 in multiple runs.

The imperfect 'cross' correlation between the tumor purity estimates by BACOM 2.0 and UNDO is expected because the two methods use different molecular data types, where copy number values are always '2' across all normal cells (e.g., stroma, T-cells, monocytes) while gene expression values are cell type specific. In fact, there are multiple gene expression profiles corresponding to various normal cells. Moreover, copy number values are generally 'static', while gene expression values are intrinsically 'dynamic'.

## 4 DISCUSSION

In this letter, we corrected and extended the BACOM method of Yu *et al.* (Yu, et al., 2011) to more accurately detect deletion types, estimate normal cell fraction, and quantify true copy numbers in tumor cells. We achieved these objectives by introducing a more comprehensive signal modeling and absolute normalization scheme (Attiyeh, et al., 2009). BACOM 2.0 offers several attractive features: (1) It performs absolute normalization by identifying the normal copy number component in a 'revised' Gaussian mixture histogram; (2) It estimates signal models and their parameter values after eliminating significant confounding factors; (3) It calculates the overall normal cell fraction (or tumor purity) with a correction for potential intratumor heterogeneity; and (4) It adjusts the effect of copy number saturation.

Fundamental to the success of our approach is the rigorous signal modeling and absolute normalization. In the presence of both normal cell contamination and tumor aneuploidy, with proper sample quality control (Popova, et al., 2009), our experience indicates that absolute normalization can be done separately (or iteratively) from tumor purity/ploidy estimation (Attiyeh, et al., 2009). We expect BACOM 2.0 to be a useful tool for analyzing copy number data in heterogeneous tumor samples (Zhang, et al., 2014), complement to the existing methods (Carter, et al., 2012; Oesper, et al., 2013).

Since BACOM 2.0 is supported by a well-grounded and unambiguous statistical framework, we foresee a variety of extensions to the concepts and strategies here. Regarding the detection of allelic-imbalanced loci, a good alternative to allelic correlation coefficient is the B allele frequency ratio (Attiyeh, et al., 2009; Popova, et al., 2009; Van Loo, et al., 2010). When there are multiple deletion segments across genome, the distribution of $\alpha$ estimates merits some further study since it may indicate the presence of intratumor heterogeneity defined by subclone copy number aberrations. Moreover, with further development, localized chromosomal ploidy can be detected instead of average tumor ploidy (Van Loo, et al., 2010).

Though a significant sample-wise correlation between the tumor purity estimates by BACOM 2.0 and ABSOLUTE has been observed, further investigation into the discrepancy between the average tumor purity estimates by the two methods would be interesting, given the fact that no definite ground truth is available. For example, TCGA used 60~80% tumor purity as the threshold to select tumor samples and a protocol baseline of 50% tumor purity was adopted to differentiate between high and low tumor purity in three cancer types (Downey, et al., 2014; Huijbers, et al., 2013; Su, et al., 2012), while rather poor correlations were reported between the estimates by ABSOLUTE/ESTIMATE and histological method probably due to miscount of infiltrating immune cells in pathological examinations (Yoshihara, et al., 2013).

*Funding*: National Institutes of Health, under Grants CA149147, CA160036, HL111362, NS29525, in part.

*Conflict of Interest*: none declared.